\documentclass[prl,twocolumn,showpacs,preprintnumbers,amsmath,amssymb,mathptmx,citeautoscript,aps,prl]{revtex4-1}    
\usepackage[dvipdfmx]{graphicx}
\usepackage{txfonts}
\usepackage{ulem}
\usepackage{cancel}

\newcommand{\fetes}{Fe$_{1+y}$Te$_{1-x}$S$_x$}
\newcommand{\FeTeS}{Fe$_{1.07}$Te$_{0.88}$S$_{0.12}$}
\newcommand{\fete}{Fe$_{1+y}$Te}
\newcommand{\fetese}{Fe$_{1+y}$Te$_{1-x}$Se$_x$}
\newcommand{\Tcon}{$T_\mathrm{c}^{\mathrm{onset}}$}
\newcommand{\Tcinf}{$T_\mathrm{c}^{\mathrm{inf}}$}
\newcommand{\Tc}{$T_\mathrm{c}$}
\newcommand{\ha}{$h_\text{anion}$}

\begin{document}

\title{Anisotropic Pressure Effects on Superconductivity in \fetes}

\author{Kazunori Yamamoto$^1$}
\author{Teruo Yamazaki$^{1,2}$}
\author{Takayoshi Yamanaka$^1$}
\author{Daichi Ueta$^3$}
\author{Hideki Yoshizawa$^3$}
\author{Hiroshi Yaguchi$^1$}

\affiliation{$^1$Department of Physics, Faculty of Science and Technology, Tokyo University of Science, Noda, Chiba 278-8510, Japan \\
$^2$Faculty of Science and Engineering, Waseda University, Shinjuku, Tokyo 169-8555, Japan \\
$^3$Institute for Solid State Physics, University of Tokyo, Kashiwa, Chiba 277-8581, Japan} 
\begin{abstract}

We have investigated the uniaxial and hydrostatic pressure effects on superconductivity 
in \FeTeS \ through magnetic susceptibility measurements down to 1.8 K.
The superconducting transition temperature \Tc \  is enhanced by out-of-plane pressure 
(uniaxial pressure along the $c$-axis); the onset temperature of the superconductivity reaches 11.8 K at 0.4 GPa.
In contrast, \Tc \ is reduced by in-plane pressure (uniaxial pressure along the $ab$-plane) and hydrostatic pressure. 
Taking into account these results, it is inferred that the superconductivity of \fetes \ is enhanced when the lattice constant $c$ considerably decreases.
This implies that the relationship between \Tc \ and the anion height for Fe$_{1+y}$Te$_{1-x}$S$_x$ is similar to that for most iron-based superconductors.
We consider the reduction of $T_\text{c}$ by hydrostatic pressure to be due to the suppression of spin fluctuations because the system moves away from antiferromagnetic ordering, and the enhancement of $T_\text{c}$ by out-of-plane pressure to be due to the anion height effect on $T_\text{c}$.
\end{abstract}


\maketitle   
\section{Introduction}
Since the discovery of superconductivity in F-doped LaFeAsO at 26 K, 
several types of iron-based superconductors have been reported \cite{Fe-based1}.
Among them, iron chalcogenides such as FeSe, \fete ,\ and \fetese \ have attracted great attention owing to their simple crystal structures, which are solely composed of 
stacked superconducting (SC) Fe$X_4$ ($X$: anion) layers \cite{various-11system}.
Therefore, iron chalcogenides are suitable for elucidating the mechanism of the superconductivity in iron-based superconductors.
Numerous experimental and theoretical studies have been carried out thus far.

Pressure experiments on superconductors are a useful technique 
for altering their electronic and magnetic properties via changes in lattice parameters.
In iron-based superconductors, many studies on pressure effects have been reported, 
and it has been revealed that the antiferromagnetic (AFM), structural, and SC transition temperatures are rather sensitive to the application of pressure.
In addition, it has been suggested that \Tc \ is correlated to the anion height \ha , which is the distance between the anion and the nearest iron layers 
\cite{Anion-height_Mizu,Anion-height_Tomi}.
The relationship between \Tc \ and \ha \ for various iron-based superconductors was summarized in a \Tc - \ha \ plot 
by Mizuguchi et al.\cite{Anion-height_Mizu}. 
They pointed out that the relationship follows a nearly symmetric curve with a peak around \ha \ $= 1.38$ \AA, 
which is the optimum value for inducing high-\Tc \ superconductivity. 
Another feature is that large pressure effects are observed in FeSe, 
whose \Tc \ greatly increases from 8 to 37 K under a pressure of 6 GPa, \cite{various-11system,FeSe-37K_Mar,FeSe-37K_Mad,FeSe-37K_Oka,FeSe-Dome_sun}. 
and the high-\Tc \ state is attributable to the vanishing of the AFM order and the enhancement of the Hall resistance related to interband spin fluctuations\cite{FeSe-Dome_sun,Sun2017PRL}.

In contrast, \fete , which is also a 11-system similar to FeSe, does not show superconductivity but undergoes AFM and structural phase transitions around 70 K.
These phase transitions are suppressed by the substitution of S for Te in \fete \ (\fetes ) or by the application of hydrostatic pressure\cite{FeTe_tra,FeTe_HP1,FeTe_HP2,FeTe_HP3,FeTe_S-sub1,FeTe_S-sub2,FeTe_S-sub3}.
In addition, superconductivity can be induced in \fetes \  
with a critical temperature of  $\sim$ 8 K 
by several kinds of post-treatments such as exposing to air\cite{air}, annealing in oxygen gas\cite{O2-1,O2-2,O2-3,O2-4}, and soaking in an organic acid\cite{acid-1,acid-2}.
It has been interpreted that such post-treatments eliminate excess iron atoms that may suppress the superconductivity in \fetes \cite{acid-2}.
Hydrostatic pressure effects on superconductivity in post-treated \fetes \ have already been investigated; \Tc \ systematically decreases with increasing pressure, in contrast to FeSe\cite{Anion-height_Mizu}.
Since \ha \ for Fe$_{1+y}$Te$_{0.9}$S$_{0.1}$ is estimated to be 1.66 \AA ,\cite{anion-height_calc} which is considerably above the optimal value of 1.38 \AA, the application of the \Tc - \ha \ relationship mentioned above to the \fetes \ system contradicts existing data obtained from hydrostatic pressure experiments\cite{Anion-height_Mizu}.
The reason for this has been an open question and we speculate that the hydrostatic pressure moves the ground state away from the AFM order and reduces AFM spin fluctuations.
In order to investigate whether the \Tc - \ha \ relationship is applicable to \fetes , a uniaxial pressure experiment is a most suitable technique.
Uniaxial pressure is expected to reduce the $c$-axis more efficiently than hydrostatic pressure.
In this study, we applied hydrostatic and uniaxial pressures parallel to the $c$-axis and $ab$-plane 
in O$_2$-annealed \fetes \ and measured the DC susceptibility using a zero-field-cooling (ZFC) procedure.
\begin{table}[htbp]
\caption{Summary of measurement conditions (HP: hydrostatic pressure, UP: uniaxial pressure). }
\label{t1}
\centering
\begin{tabular}{lllll} 
\hline
\multicolumn{1}{c}{Pressure} & \multicolumn{1}{c}{Cell}  & \multicolumn{1}{c}{Medium} & \multicolumn{1}{c}{Dimensions}\\
\hline \hline
HP & cell 1 & Daphne 7373 oil & $\phi1.2 \times 0.7$ mm$^3$ \\
UP [001] direction & cell 2 & not used & $\phi 1.6 \times 0.4$ mm$^3$ \\
UP [100] direction & cell 3 & not used & $1.3 \times 1.2 \times 1.2$ mm$^3$ \\
UP [110] direction & cell 3 & not used & $1.3 \times 1.2 \times 1.2$ mm$^3$ \\
\hline
\label{table1}
\end{tabular}
\end{table}

\section{Experimental Procedure}
\begin{figure}[htbp]
\centering 
\vspace{0cm}
\includegraphics[width=1.0\linewidth]{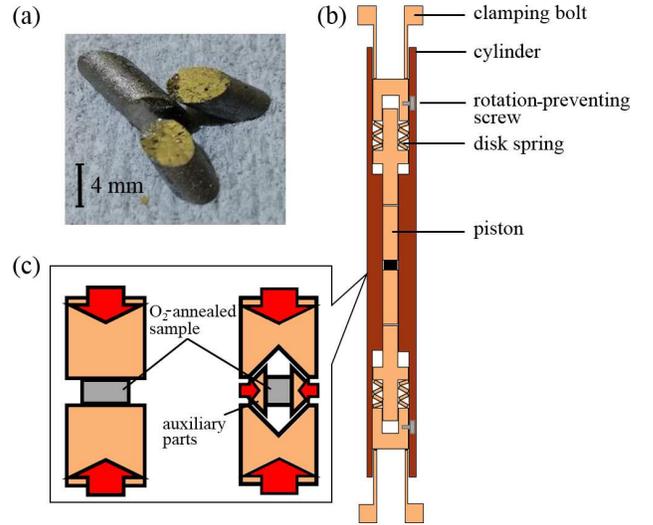}
\caption{(Color online) (a) Photograph of the single crystals of \FeTeS \ grown in this study. Schematic diagrams of (b) uniaxial pressure cell and (c) enlarged view of the pistons and sample. }
\label{pressure-cell}
\end{figure}

The single-crystalline samples of \fetes \ used in this study were prepared by Tammann's method.
Fe cube (5N), Te shot (6N), and S shot (6N) with a nominal composition of $x$ = 0.2, $y$ = 0 were sealed in a quartz ampule under an atmosphere of 0.3 atm argon.
The ampule was heated at 1050 $^{\circ}$C for 20 h by an electric furnace in accordance with Refs. 16 and 25.
To refine the crystals, the obtained samples were ground and sealed in a quartz ampule filled with 0.3 atm of argon gas.
The ampule was set in the furnace with a temperature gradient of 3 $^{\circ}$C/cm, heated to 950 $^{\circ}$C in 8 h, kept at 950 $^{\circ}$C for 20 h, cooled down to 650 $^{\circ}$C at a rate of $3$ $^{\circ}$C/h and kept at 650 $^{\circ}$C for 20 h.
Large single crystals ($\phi $4 $\times$ 30 mm$^3$) were obtained [Fig. \ref{pressure-cell}(a)].
For sample characterization, powder X-ray diffraction patterns of the grown samples were obtained by a Rigaku Ultima I\hspace{-.1em}V diffractometer.
From the results, we confirmed that the samples were of single phase without any apparent component peak from impurities such as an FeTe$_2$.
The actual composition of the crystal was determined to be \FeTeS \ using an electron probe microanalyzer (EPMA; JEOL JXA-8100).
Samples used for pressure measurements were shaped by wire electrical discharge machining as tabulated in Table \ref{table1}. 
To induce superconductivity, these samples were annealed in atmospheric O$_2$ gas at 200 $^{\circ}$C for 100 h.
Then, the crystallographic orientation was determined by the Laue method with OrientExpress simulation software.
Electrical resistivity measurements were performed using a four-terminal method from room temperature to 4.2 K.
The specific heat was measured by a thermal relaxation method (Quantum Design, PPMS).
Magnetic susceptibility measurements under ambient, hydrostatic, and uniaxial pressures were performed down to 1.8 K using a superconducting quantum interference device (SQUID) magnetometer (Quantum Design, MPMS).
A magnetic field ($H$) of 20 Oe was applied parallel to the $c$-axis in the ZFC sequence.

Hydrostatic and uniaxial pressures were applied using clamped piston-cylinder-type pressure cells.
In this study, three types of pressure cells were used to apply hydrostatic pressure (denoted as ``cell 1'') and uniaxial pressure along the out-of-plane (denoted as ``cell 2'') and in-plane (denoted as ``cell 3'') directions.
In these cells, samples were placed such that the $c$-axis was along the cylinder, and hence the magnetic field was parallel to the $c$-axis.
For cell 1 the cylinder was made of BeCu and the pistons were made of ZrO$_2$.
Daphne 7373 oil was used as the pressure-transmitting medium.
The actual pressure was estimated from the SC transition temperature of a Pb manometer. 
Cell 2 and cell 3 were made of BeCu [Fig. \ref{pressure-cell}(b),(c)].
In cell 2 [Fig. \ref{pressure-cell}(c) left], the sample was directly pressed along the $c$-axis by the pistons.
On the other hand, in cell 3 [Fig. \ref{pressure-cell}(c) right] the sample was kept in contact with the pair of auxiliary parts [as illustrated in Fig. \ref{pressure-cell}(c) right] and pressed horizontally  when forces were vertically applied on the V-shaped pistons.
Consequently, a uniaxial pressure along the $ab$-plane was realized under $H \parallel c$ within the narrow sample space of the MPMS.
The side surfaces of the sample used for uniaxial pressure measurements were covered with epoxy resin (Stycast 1266) to prevent the collapse of the sample. 

\section{Experimental Results}
\begin{figure}[htbp]
\centering 
\includegraphics[width=1.0\linewidth]{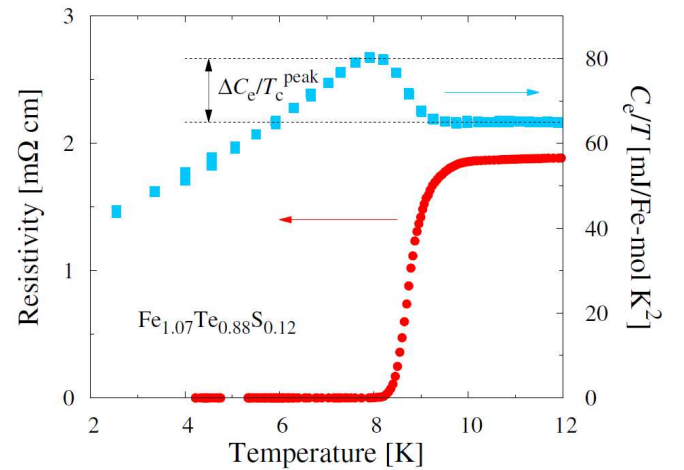}
\caption{(Color online) Temperature dependence of electrical resistivity and electronic specific heat divided by temperature $C_{\mathrm{e}}$/$T$ in \FeTeS \ at ambient pressure.} 
\label{RT-CT}
\end{figure}

Figure \ref{RT-CT} shows the temperature dependence of the resistivity and specific heat of O$_2$-annealed \FeTeS \ at ambient pressure.
We observed a clear drop in the resistivity and a discontinuity in the specific heat, which are attributable to an SC transition.
The resistivity starts dropping at a temperature above 9 K and reaches zero at 8.2 K.
The electronic specific heat divided by the temperature was obtained by subtracting the lattice contribution ($C_\text{ph} \propto T^3$), and is plotted on the right axis in Fig. \ref{RT-CT}.
A jump associated with the SC transition was observed at $T_{\mathrm c}^{\mathrm {peak}}$ = 8 K.
Then the SC volume fraction was roughly estimated to be 16\% by assuming the BCS theory and using the values of $\Delta C_{\mathrm e}/T_{\mathrm c}^{\mathrm {peak}}$ and the electronic specific heat coefficient $\gamma$, where $\Delta C_{\mathrm e}/T_{\mathrm c} = 1.43\gamma $ for the BCS theory.
The small SC volume fraction of 16\% is due to the O$_2$-annealing effect penetrating the sample over a depth of 10 $\mu $m orders\cite{size}. 

\begin{figure}[htbp]
\centering 
\includegraphics[height=22.5cm]{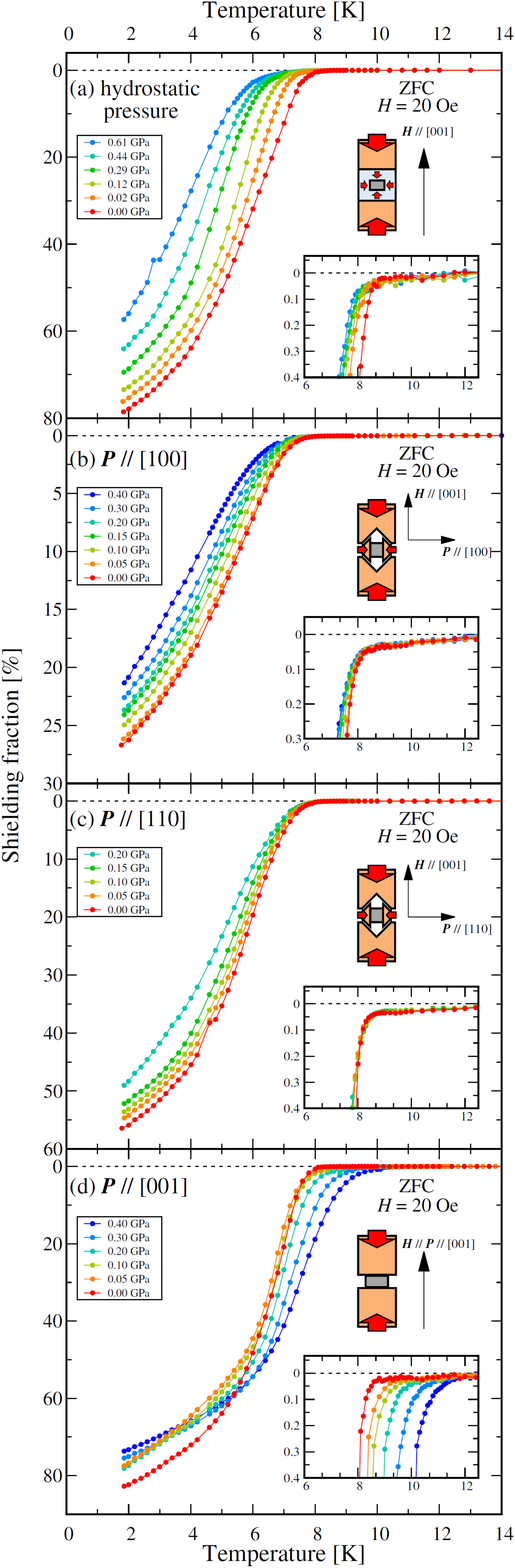}
\caption{(Color online) Temperature dependences of Meissner shielding fraction in \FeTeS \ obtained from susceptibility measurements under (a) hydrostatic pressure and under uniaxial pressure along the (b) [100], (c) [110], and (d) [001] directions.
The insets show enlargements around the starting point of the drop in the magnetic susceptibility.}
\label{TvsSF}
\end{figure}

Figure \ref{TvsSF} shows the temperature dependence of the Meissner shielding fraction (SF) in O$_2$-annealed \FeTeS \ under (a) hydrostatic pressure and under uniaxial pressure along the (b) [100], (c) [110], and (d) [001] directions. 
The demagnetizing factor used was based on the approximation of the sample shape by a spheroid. 
The insets show enlargements around the starting point of the drop corresponding to the onset of the SC transition.
Here we use two definitions of the SC transition temperature.
One is the onset of the drop in magnetization, denoted as \Tcon .
The other is the point of inflection associated with the SC transition, denoted as \Tcinf .
At ambient pressure, \Tcon \ is determined to be about 8.4 K for all of the samples.

\begin{figure}[htbp]
\centering 
\includegraphics[width=1.0\linewidth]{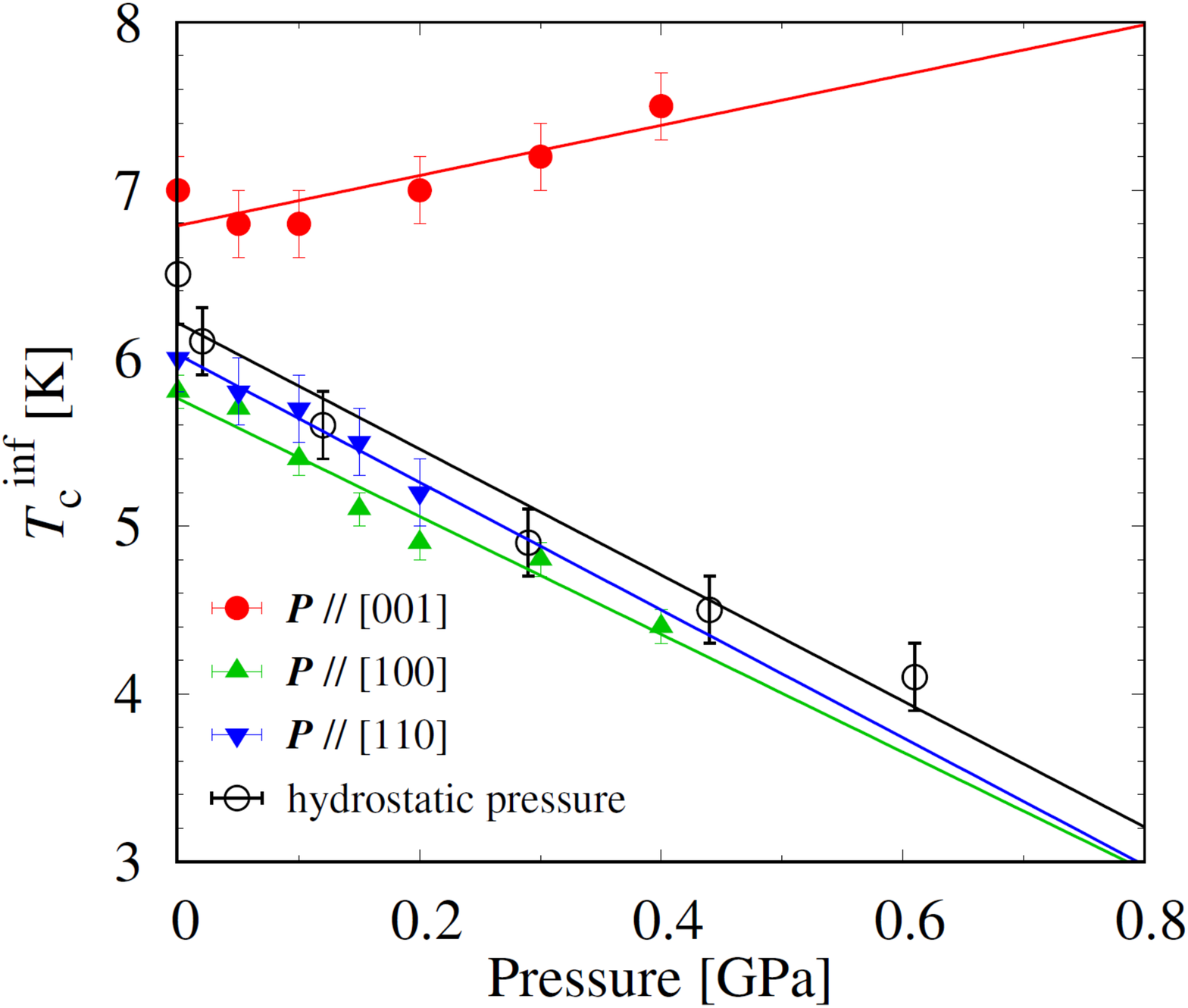}
\caption{(Color online) Hydrostatic and uniaxial pressure dependences of \Tcinf \ in \FeTeS . }
\label{PTc}
\end{figure}

Under hydrostatic pressure, both \Tcon \ and \Tcinf \ decrease, yielding a negative value of $-3.8 \pm 0.4$ K/GPa for $\frac{\mathrm{d} T_{\text c}^{\text {inf}}}{{\mathrm d} P}$, in reasonable agreement with Ref. 3.
Under uniaxial pressure along the [100] and [110] directions, \Tcinf \ and the SF at 1.8 K systematically decrease with increasing pressure.
\Tcon \ hardly changes with increasing pressure.
This is due to the broadening of the SC transition caused by the inhomogeneity of the applied pressure.
The region under the lowest pressure is probably responsible for \Tcon .
For out-of-plane pressure, \Tcon \ systematically increases [inset of Fig. \ref{TvsSF}(d)] and reaches 11.8 K at 0.4 GPa. 
\Tcinf \ also tends to increase above 0.1 GPa. 
Thus, uniaxial pressure along the $c$-axis enhances \Tc \ while \Tc \ is reduced by uniaxial pressure along the $ab$-plane, similar to hydrostatic pressure.
We plot the relationship between pressure and \Tc \ in Fig. \ref{PTc}, where we employ \Tcinf \ to evaluate the average effects of uniaxial pressure over the whole sample.
We have determined the pressure coefficient of the SC transition temperature $\frac{\text{d}T_\text{c}}{\text{d}P}$ by linear fitting as shown in Fig. \ref{PTc}. 
Those for uniaxial pressure along the [100], [110], and [001] directions are $-3.5 \pm 0.3$, $-3.8 \pm 0.4$, and $1.5 \pm 0.5$ K/GPa, respectively.
Since the pressure coefficients under uniaxial pressure along the [100] direction and [110] direction are almost the same, the in-plane anisotropy of the pressure appears to be little.
However, the anisotropy between in-plane and out-of-plane pressure is prominent; the pressure coefficient is different in sign.

\section{Discussion}
According to hydrostatic pressure studies on FeSe\cite{FeSe-Dome_sun}, \Tc \ reaches 37 K  at 6 GPa with the coincidental vanishing of AFM order, and then \Tc \ is gradually reduced at higher pressure than 6 GPa.
A recent study reported that the suppression of $T_\text{c}$ is concomitant with the reduction of the AFM fluctuations \cite{Sun2017PRL}. 
In the case of \fetes , sulfur substitution with the composition of $x = 0.06 - 0.12$ suppresses the AFM order and structural transition \cite{O2-3}, and the application of hydrostatic pressure decreases $T_\text{c}$. 
Thus, the $T$-$P$ phase diagram of \FeTeS \ is simpler than that of FeSe and they are very different.
The suppression of \Tc \ in \FeTeS \ by the application of hydrostatic pressure is likely to originate from the reduction of AFM fluctuations by moving away from the AFM phase, which is similar to the case of FeSe under hydrostatic pressure ($>$ 6 GPa).
In contrast, the application of uniaxial pressure along the $c$-axis increases \Tc \ in \FeTeS .
Straightforwardly, the uniaxial pressure along the $c$-axis reduces \ha \ more efficiently than hydrostatic pressure, and the enhancement of \Tc \ is considered to be due to \ha \  approaching the optimum value.
The applicability of the \ha - \Tc \ relationship to \fetes \ is supported by the anisotropic pressure effects on \Tc .
From these results, it is expected that the effect of moving away from the AFM phase competes with the effect of changing \ha \ toward the optimum value in the hydrostatic pressure case.
However, in order to determine whether this expectation is correct or not, it is necessary to measure the spin fluctuations under uniaxial and hydrostatic pressures by magnetic and dynamic probes, such as the relaxation time of nuclear magnetic resonance.

\section{Conclusions}
We have investigated the pressure effects on superconductivity in \FeTeS \ by susceptibility measurements under hydrostatic pressure and uniaxial pressure along the [001], [100], and [110] directions.
We observed the enhancement of \Tc \ by uniaxial pressure along the $c$-axis, which probably shrinks \ha .
In contrast, \Tc \ is reduced by uniaxial pressure along the $ab$-plane.
Although it has been considered that the relation between \ha \ and \Tc \ is not applicable to \fetes \cite{Anion-height_Mizu},
 our results suggest that $c$-axis uniaxial pressure greatly reduces the anion height and increases \Tc \ while obeying the \ha -\Tc \ relationship.
From the hydrostatic and $c$-axis uniaxial pressure studies, \ha \ might be independent of AFM spin fluctuations, but it is necessary to obtain magnetic information by magnetic and dynamical probes.
Finally, we consider \FeTeS \ as a suitable research field on the \ha -\Tc \ relationship because of the simple $T$-$P$ phase diagram.

\section*{Acknowledgments}
This work was partly carried out under the Visiting Researcher$^\prime$s Program of the Institute for Solid State Physics (ISSP), the University of Tokyo.
We would like to thank Y. Shimizu, Tokyo University of Science, for his help in the resistivity measurements.


\end{document}